\documentclass[12pt, a4paper]{article}
\usepackage{graphicx}

\newcommand{\be}{\begin{equation}}
\newcommand{\ee}{\end{equation}}
\newcommand{\bea}{\begin{eqnarray}}
\newcommand{\eea}{\end{eqnarray}}
\newcommand{\bdm}{\begin{displaymath}}
\newcommand{\edm}{\end{displaymath}}

\newcommand{\p}{\partial}
\newcommand{\curl}{\mbox{curl}}
\newcommand{\dive}{\mbox{div}}
\newcommand{\const}{\mbox{const.}}
\newcommand{\diff}{\mbox{$\rm{d}$}}

\newcommand{\bbg}{\mbox{\boldmath $g$}}
\newcommand{\bbx}{\mbox{\boldmath $x$}}
\newcommand{\bbv}{\mbox{\boldmath $v$}}
\newcommand{\bbz}{\mbox{\boldmath $z$}}
\newcommand{\bbA}{\mbox{\boldmath $A$}}
\newcommand{\bbB}{\mbox{\boldmath $B$}}
\newcommand{\bbE}{\mbox{\boldmath $E$}}
\newcommand{\bbJ}{\mbox{\boldmath $J$}}
\newcommand{\bbS}{\mbox{\boldmath $S$}}
\newcommand{\bbmu}{\mbox{\boldmath $\mu$}}
\newcommand{\bbomega}{\mbox{\boldmath $\omega$}}
\newcommand{\bbnabla}{\mbox{\boldmath $\nabla$}}

\def\Box{\hbox{\hskip 0.5mm\hbox{\vrule width2.3mm height0.2mm
\vbox{\hrule width0.3mm height2.6mm}\hskip
-2.6mm \vbox{\hbox{\vrule width2.6mm height0.1mm}
\vskip -0.1mm\hrule width0.1mm height2.6mm}}\hskip 0.5mm}}

\begin{document}

\title{Reflections on Gravity\footnote{
Concluding talk at the {\bf ESA-CERN WORKSHOP}, CERN, 5-7 April 2000}
}

\author{Norbert Straumann\\
        Institute for Theoretical Physics, University of Zurich,\\
        CH--8057 Zurich, Switzerland}
\date{\today}

\maketitle

\begin{abstract}
A pedagogical description of a simple ungeometrical approach
to General Relativity is given, which follows the pattern of well
understood field theories, such as electrodynamics. This leads
quickly to most of the important weak field predictions, as
well as to the radiation damping of binary pulsars. Moreover,
certain consistency arguments imply that the theory has to
be generally invariant, and therefore one is bound to end
up with Einstein's field equations. Although this field theoretic
approach, which has been advocated repeatedly by a number
of authors, starts with a spin-$2$ theory on Minkowski spacetime,
it turns out in the end that the flat metric is
actually unobservable, and that the physical metric is
curved and dynamical.

Short sections are devoted to tensor-scalar generalizations,
the mystery of the vacuum energy density, and quintessence.
\end{abstract}

\section{Introductory Remarks}

I feel very honored by the invitation to give the concluding
talk at this exciting workshop. At the same time I feel
a bit worried. After the excellent summaries of the various
parallel sessions I shall, of course, not give another
overview. Before indicating what I will concentrate on,
let me begin with a few general remarks.

At the time in 64/65 when I was a Fellow here in the CERN theory
group, General Relativity (GR) played virtually no role in 
high energy physics and was even largely unknown to
particle physicists. And this remained so for quite some
time, essentially until the advent of supergravity theory.

With the unification attempts in the framework of
supergravity theory, the revival of (supersymmetric)
Kaluza-Klein theories, and finally string \mbox{(M-)} theory,
gravitational interactions became an {\it essential} and
{\it unavoidable} part of fundamental (speculative) physics.

Another reason - more familiar to the participants of this workshop -
why gravitational physics has become a central field in present
day physics, is the incredible chain of important astronomical
discoveries since the early sixties. We are undoubtedly
in the middle of a truly Golden Age of astrophysics and
cosmology. Soon we shall have a gravitational wave astronomy,
allowing us to study highly dynamical strong field processes,
like the coalescence of black holes. Here Einstein's equations
come into play in their full glory. Surely, gravitational
wave searches will also transform GR into a field like other
branches of physics, with a healthy interplay of theory and
experiment. For the analysis of the expected data, a lot
of difficult analytical and numerical work remains to be done.

In this situation every physicist should have some
technical understanding of this marvellous relativistic
theory of gravity, called GR, about which Dirac once
said that it ``is probably the greatest scientific discovery
that was ever made''.

An obstacle for a full understanding of GR has
always been the necessity of absorbing first a considerable
amount of mathematical machinery. This is, of course,
no problem for theoreticians, but experimentalists and
astronomers often do not find the time for this.
(This is at least true for many people I know.)
To some extend, this hurdle can be postponed in an
ungeometrical approach to GR, which has been advocated
in the course of time by a number of authors, in
particular by R.P. Feynman in his Caltech lectures \cite{1}.
One may call it the flat spacetime - or the field
theoretic approach. I first learned about it in my
youth from discussions with M. Fierz. That was shortly
after he left CERN as theory director and came to Zurich
as Pauli's successor. His ideas were partially worked
out in the thesis of W. Wyss \cite{2}. At about the same
time W. Thirring was advocating this approach with
different emphasis in talks and some publications \cite{3}.
S. Weinberg had a related paper \cite{4}, in which he made an
attempt to develop a quantum theory of a selfinteracting
spin-$2$ field on flat spacetime. (We now know that
such theories are unrenormalizable, also for supersymmetric
extensions.) The theme was taken up later by S. Deser \cite{5},
R.M. Wald \cite{6}, and others.

The idea of this alternative approach is to describe
gravity - in close analogy to electrodynamics - by a
field theory on {\it flat} Minkowski spacetime. I shall
spend much of my time in showing you how this
can be done without much effort. One of the interesting
lessons of this will be the following: although we shall
start with a theory on Minkowski spacetime, it will in the end
turn out that the flat metric is actually {\it unobservable},
and that the physical metric is {\it curved} and {\it dynamical},
subject to field equations which agree with Einstein's
equations. The physical metric is determined by the
spin-$2$ field of the theory. In other words, the initial
flat metric turns out to be a kind of unobservable ether
which we will eliminate.

One of the advantages of this field theoretic
approach is that it follows the patterns of well understood
field theories and may thus be closer to what many of
you are used to. Other pros and cons will be discussed
later.

To some of you, most of what I am going to say is not new,
but I hope that perhaps a majority of the participants of
this workshop will afterwards see GR in a somewhat
different light. This can also be useful as a starting
point for discussing experimental tests. However,
since this subject has been thoroughly reviewed by
other speakers, I shall confine myself to a few
scattered remarks, mostly in connection with tensor-scalar
generalizations of GR. I shall end with some comments on
the $\Lambda$-problem.

\section{A field theoretic (pedestrian) approach to GR}

A natural question shortly after 1905 was: Why not
develop a field theory of gravity in close analogy to
electrodynamics? Einstein's first (unpublished)
attempts went in this direction.

In trying to do this, it is useful to recall the
following avenue to Maxwell's equations.

As starting point we adopt the following three
ingredients:
\begin{enumerate}
\item[(i)]{{\bf Electrostatics:}
\bea
&& \bullet\;\hbox{field equations:}\;\;\; \curl\bbE = 0, \;\;\; \dive\bbE = \rho_e
   \label{Eq-1}\\
&& \bullet\;\hbox{equation of motion for test particle:}\;\;\; m\ddot{\bbx} = e\bbE.
   \label{Eq-2}
\eea
}
\item[(ii)]{{\bf Lorentz-invariance} of the theory. Today this is a
battle-field-tested general symmetry principle.
[ Historically, it grew out of electrodynamics,
but now we have overwhelming direct experimental
evidence for its validity. I do not have to
stress this here at CERN.]
}
\item[(iii)]{{\bf Charge conservation:}
\be
\p_\mu j^{\mu} = 0, \;\;\; j^{\mu} = (\rho_e, \bbJ ).
\label{Eq-3}
\ee
[ This is presumably the most important and most general
conservation law of physics.]
}
\end{enumerate}
On the basis of (i)-(iii), Maxwell's equations are more than
compelling. We argue as follows:

The source of the electromagnetic field must be the current $j^{\mu}$,
instead of $\rho_e$ in (\ref{Eq-1}). The Lorentz-invariant generalization
of the inhomogeneous equation in (\ref{Eq-1}) will be of the symbolic
form $(\p\cdot F)^{\mu} = -j^{\mu}$, where $\p\cdot F$ denotes the divergence
of a tensor field. Clearly, $F$ has to be a tensor field
of second rank, and the inhomogeneous field equation is
naturally expected to be of the form $\p_\nu F^{\mu\nu} = -j^{\mu}$.

Now, following Maxwell, we require that current conservation
is an automatic consequence of the field equations. This
can only be the case if $F^{\mu\nu}$ is antisymmetric. The
field tensor then transforms {\it irreducibly} with respect to
the homogeneous Lorentz group.
With similar arguments one is also led to the homogeneous
Maxwell equations. Note that Lorentz-invariance implies, in
particular, the existence of the magnetic field for
nonstatic situations. Similarly, we shall predict in
gravity theory the existence of a gravitomagnetic field.

This compelling reasoning of guessing the correct
relativistic field equations is now taken as a model
for gravity. We start again from three similar ingredients:
\begin{enumerate}
\item[(i)]{{\bf Static limit} (Newtonian theory): The gravitoelectric
field $\bbg$ (gravitational acceleration of test bodies) satisfies
the field equations:
\be
\curl\,\bbg = 0, \;\;\; \dive\,\bbg = -4\pi G\rho
\label{Eq-7}
\ee
or in terms of the potential $\varphi$,
\be
\bbg = -\bbnabla\varphi, \;\;\; \Delta\varphi = 4\pi G\rho.
\label{Eq-8}
\ee
The equation of motion for a test body is universal (weak
equivalence principle):
\be
\ddot{\bbx} = \bbg(\bbx), \;\;\; (m_i = m_g).
\label{Eq-9}
\ee
}
\item[(ii)]{{\bf Lorentz-invariance};}
\item[(iii)]{{\bf Energy-momentum conservation:}
\be
\p_\nu T^{\mu\nu} = 0.
\label{Eq-10}
\ee
}
\end{enumerate}
Since this is so similar to ED, one expects at first
sight that it should not be difficult to find a
satisfactory Lorentz-invariant theory of gravity. Let us
try this.

First, we have to find out which type of field has
to be chosen. A priori, we have the possibilities
$0$, $1$ or $2$ for the spin.
Among these the spin-$1$ option has to be excluded,
because it would lead to {\it repulsion}. That this is
unavoidable was already known to Maxwell. Otherwise,
the gravitational waves would have negative energy and
the world would be unstable. (This will soon
become clear.)

\subsection{Scalar theory}

It is instructive to consider first the simplest case
of a scalar theory (Einstein, Nordstr\"om, v. Laue \cite{7}).
(I do this also because scalar-tensor theories are
still under current discussion, see section \ref{Sect-3}.)

The field equation for the scalar field $\varphi$ in the
{\it weak} field case (linear field equation) is unique:
\be
\Box\varphi = -4\pi G\, T, \;\;\; T:=T^{\mu}_{\;\mu}.
\label{Eq-11}
\ee
(For a static situation this reduces to the second equation
in (\ref{Eq-8}).)

We formulate the equation of motion of a test particle
in terms of a Lagrangian. For weak fields this is again
unique:
\be
L(x^{\mu}, \dot{x}^{\mu} ) = -\sqrt{\eta_{\mu\nu} \dot{x}^{\mu}\dot{x}^{\nu}} (1 + \varphi),
\label{Eq-12}
\ee
because only for this the Newtonian limit for weak
static fields and small velocities of the test bodies
comes out right:
\bdm
L( \bbx, \dot{\bbx} ) \approx \frac{1}{2} \dot{\bbx}^2 - \varphi + \const
\edm

The basic equations (\ref{Eq-11}) and (\ref{Eq-12}) imply a perihelion
motion of the planets, but this comes out wrong, even
the sign is incorrect. One finds (-1/6) times the value of GR.
In spite of this failure I add some further instructive remarks.

First, I want to emphasize that the interaction is
necessarily {\it attractive}, independent of the matter content. To
show this, we start from the general form of the Lagrangian
density for the scalar theory
\be
{\cal L} = \frac{1}{2} \p_\mu\phi\, \p^\mu \phi - g\, T\cdot\phi + {\cal L}_{mat}.
\label{Eq-13}
\ee
($\phi$ is proportional to $\varphi$; $g$ is a coupling constant.)

Note first that only $g^2$ is significant: Setting
$\tilde{\phi} = g\phi$, we have
\bdm
{\cal L} = \frac{1}{2g^2} \p_\mu\tilde{\phi}\, \p^\mu \tilde{\phi} - T\cdot\tilde{\phi} + {\cal L}_{mat},
\edm
involving only $g^2$. Next, it has to be emphasized that it
is not allowed to replace $g^2$ by $-g^2$, otherwise the
field energy of the gravitational field would be negative.
(This ``solution'' of the energy problem does not work.)
Finally, we consider the field energy for {\it static} sources.

The total (canonical) energy-momentum tensor
\bdm
T^\mu_{\;\nu} = \frac{\p {\cal L}}{\p \phi_{,\mu}} \phi_{,\nu} + ... - \delta^{\mu}_{\;\nu} {\cal L}
\edm
gives for the $\phi$-contribution:
\bdm
(T_\phi)_{\mu\nu} = \p_\mu\phi\, \p_\nu\phi - \frac{1}{2} \eta_{\mu\nu} \p_\lambda\phi\, \p^{\lambda}\phi
 + \eta_{\mu\nu} g\, T\phi .
\edm
For the corresponding total energy we find
\bea
E &=& \int (T_\phi)_{00} \diff^3 x = \frac{1}{2} \int \left[ (\bbnabla \phi)^2 + 2 g\, T\phi \right] \diff^3 x
\nonumber\\
  &=& \frac{1}{2} \int \left[ \phi\,(-\Delta\phi) + 2 g\, T\phi \right] \diff^3 x
   =  \frac{1}{2}\, g \int T\phi\, \diff^3 x. \nonumber
\eea
Since $\Delta\phi = g\, T$, we have
\bdm
\phi(\bbx) = -\frac{g}{4\pi} \int \frac{ T(\bbx') }{ |\bbx - \bbx'| }\, \diff^3 x'.
\edm
Inserting this above gives finally
\bdm
E = -\frac{g^2}{4\pi} \frac{1}{2} \int \frac{ T(\bbx) T(\bbx') }{ |\bbx - \bbx'| } \diff^3 x\, \diff^3 x',
\edm
showing that indeed the interaction is attractive.

This can also be worked out in quantum field theory
by computing the effective potential corresponding to
the one-particle exchange diagram with the interaction
Lagrangian ${\cal L}_{int} = g\, \bar{\psi} \psi\, \phi_{m=0}$.
One finds
\bdm
V_{eff} = -\frac{g^2}{4\pi} \frac{1}{|\bbx - \bbx'|},
\edm
both for fermion-fermion and fermion-antifermion interactions.
The same result is found for the exchange of massless spin-$2$ particles,
while for spin-$1$ we obtain {\it repulsion} between particles,
and attraction between particles and antiparticles.

The scalar theory predicted that there is {\it no light deflection},
simply because the trace of the electromagnetic
energy-momentum tensor vanishes. For this reason
Einstein urged in 1913 astronomers (Erwin Freundlich
in Potsdam) to measure the light deflection during
the solar eclipse the coming year in the Krim. Shortly
before the event the first world war broke out. Over
night Freundlich and his German colleagues were
captured as prisoners of war and it took another five
years before the light deflection was observed.

Quite independently of the failures mentioned so far,
an even more profound difficulty of the scalar theory is
often claimed. The argument is based on the observation
that the trace $T(\bbx, t)$ for a moving particle is given
by
\bdm
T(\bbx, t) = m \sqrt{ 1 - \bbv^2 }\, \delta^3 (\bbx - \bbz(t) ),
\edm
where $\bbz(t)$ is the position of the particle, $\bbv = \dot{\bbz}$.
The momentum dependence shows that the moving particle
generates a {\it weaker} gravitational field than a particle
at rest.

Now, consider two boxes each containing $N$ particles
of mass $m$, initially at rest (gas at zero temperature).
We imagine that the two containers 1 and 2 are connected
by a rigid rod. The two boxes gravitationally
attract each other. We can arrange things such that the
forces balance each other and that the system is at rest.
Suppose now that the rest mass of a single
particle in box 2 is completely transformed into kinetic
energy of the remaining particles. This does not change
the inertial mass of container 2, but apparently the
{\it active} gravitational mass becomes smaller, and the
total system begins to accelerate. Terrible!

There is a subtle error in this argument. The moving
particles are bouncing on the wall of box 2. This induces
a surface tension, generating an additional gravitational
field. It is easy to show that it is because of this
that the total system remains at rest.

This nicely illustrates that the equivalence principle
is a subtle and profound property of gravity.

Some final remarks on the scalar theory. So far we
have only considered weak fields generated by $T_{matter}$.
It is however, more than natural that the energy-momentum
tensor of the gravitational field also acts as a source,
so that the theory has to be nonlinear. The nonlinear
generalization of Nordstr\"om's theory was set up by
Einstein and Fokker in 1914 \cite{8}. In a non-geometrical
(flat-spacetime) formulation the Lagrangian is given by
\bdm
{\cal L} = \frac{1}{2} \p_\mu\varphi\, \p^\mu\varphi
         + {\cal L}_{mat}\left[ \psi; (1 + \kappa\varphi)^2\eta_{\mu\nu} \right] (1 + \kappa\varphi)^4;
\edm
in particular, the flat metric $\eta_{\mu\nu}$ in ${\cal L}_{mat}$ is replaced by
$(1 + \kappa\varphi)^2 \eta_{\mu\nu}$, $\kappa^2 = 8\pi G$.

One can eliminate the Minkowski metric and replace
it by a ``physical metric'':
\bdm
g_{\mu\nu} = (1 + \kappa\varphi)^2 \eta_{\mu\nu}\, .
\edm
For example, only relative to this metric the Compton wave
length is constant, i.e., not spacetime dependent.

Einstein and Fokker gave a geometrical formulation
of the theory. This can be summarized as follows:
\begin{enumerate}
\item[(i)]{ spacetime is conformally flat: Weyl tensor $=0$;}
\item[(ii)]{ field equation: $R = 24\pi G\, T$;}
\item[(iii)]{ test particles follow geodesics.}
\end{enumerate}
In adapted coordinates, with $g_{\mu\nu} = \phi^2 \eta_{\mu\nu}$,
one finds
\bdm
R = -6\eta^{\mu\nu} \p_\mu\phi\, \p_\nu\phi / \phi^3,
\edm
and the field equation becomes
\bdm
\eta^{\mu\nu} \p_\mu\p_\nu\phi = -4\pi G \phi^3\, T.
\edm

The Einstein-Fokker theory is generally covariant (as
emphasized in the original paper), however, {\it not} generally
{\it invariant}. I use this opportunity to point out the
crucial difference of the two concepts. For a long time
people (including Einstein) were not fully aware of this
and this caused lots of confusion and strange controversies.
(See, e.g., the preface of Fock's book on GR.)

The {\it invariance} group of a theory is the subgroup of the
covariance group that leaves the absolute, non-dynamical
elements of the theory invariant. In the Einstein-Fokker
theory the conformal structure is an {\it absolute} element,
and therefore the invariance group is the {\it conformal group},
whence a finite dimensional Lie group. In GR,
on the other hand, the metric is entirely dynamical,
and therefore the covariance group is at the same time
also the invariance group. For this reason, ``general relativity''
is an appropriate naming, never mind Fock and others.

A very remarkable property of the Einstein-Fokker
theory is, that it satisfies even the {\it strong} equivalence
principle. Beside GR this is (to my knowledge) the only
theory which does this. The Einstein-Fokker theory shows,
and this is a bit puzzling, that the equivalence principle
does {\it not} imply light deflection.
(For further discussion of this, see \cite{9}.)

This was probably too much on the scalar theory, but I wanted to
make some points of general significance which can more easily
be explained in this context. Let us now turn to the tensor theory.

\subsection{Tensor (spin-$2$) theory}

We are led to study the spin-$2$ option. (There are no
consistent higher spin equations with interaction.) This
means that we try to describe the gravitational field by
a symmetric tensor field $h_{\mu\nu}$.

Such a field has $10$ components. On the other hand,
we learned from Wigner that in the massless case there
are only {\it two} degrees of freedom.
How do we achieve the truncation from $10$ tow $2$ ?

Recall first the situation in the {\it massive} case. There
we can require that the trace $h = h^{\mu}_{\;\mu}$ vanishes,
and then the field $h_{\mu\nu}$ transforms with respect to the
homogeneous Lorentz group irreducibly as $D^{(1,1)}$
(in standard notation).
With respect to the subgroup of rotations this reduces to the
reducible representation
\bdm
D^1 \otimes D^1 = D^2 \oplus D^1 \oplus D^0.
\edm
The corresponding unwanted spin-$1$ and spin-$0$ components
are then eliminated by imposing $4$ subsidery conditions:
\bdm
\p_\mu h^\mu_{\;\nu} = 0.
\edm
The remaining $5$ degrees of freedom describe (after quantisation)
massive spin-$2$ particles (Pauli and Fierz \cite{10}; see, e.g.,
the classical book of G. Wentzel \cite{11}).

In the {\it massless} case we have to declare certain
classes of fields as physically equivalent, by imposing
- as in ED - a {\it gauge invariance}. The gauge
transformations are
\be
h_{\mu\nu} \longrightarrow h_{\mu\nu} + \p_\mu\xi_\nu + \p_\nu\xi_\mu,
\ee
where $\xi_\mu$ is an arbitrary vector field.

Let us first consider the {\it free} spin-$2$ theory which is
unique (Pauli and Fierz):
\be
{\cal L} = \frac{1}{4} h_{\mu\nu, \sigma} h^{\mu\nu, \sigma}
         - \frac{1}{2} h_{\mu\nu, \sigma} h^{\sigma\nu, \mu}
         - \frac{1}{4} h_{,\sigma} h^{,\sigma}
         - \frac{1}{2} h_{,\sigma} h^{\nu\sigma}_{\;\;\; ,\nu}\, .
\label{Eq-15}
\ee
Let $G_{\mu\nu}$ denote the Euler-Lagrange derivative of ${\cal L}$
(up to a sign),
\bea
G_{\mu\nu} &=& \frac{1}{2} \p^{\sigma} \p_{\sigma} h_{\mu\nu} + \p_\mu \p_\nu h
            - \p_\nu \p^\sigma h_{\mu\sigma} - \p_\mu \p^\sigma h_{\sigma\nu} \nonumber\\
           &+& \eta_{\mu\nu} \left( \p^\alpha \p^\beta h_{\alpha\beta} - \p^\sigma \p_\sigma h \right). 
\label{Eq-16}
\eea
The free field equations
\be
G_{\mu\nu} = 0
\label{Eq-17}
\ee
are identical to the linearized Einstein equations and describe,
for instance, the propagation of weak gravitational fields.

The gauge invariance of ${\cal L}$ (modulo a divergence)
implies the identity
\be
\p_\nu G^{\mu\nu} \equiv 0
\label{Eq-18}
\ee
(``linearized Bianchi identity'').
[ This should be regarded in analogy to the identity
$\p_\mu( \Box A^\mu - \p^\mu\p_\nu A^\nu) \equiv 0$
for the left-hand side of Maxwell's equations.]

Let us now introduce couplings to matter. The simplest
possibility is the linear coupling
\be
{\cal L}_{int} = -\frac{1}{2} \kappa\, h_{\mu\nu} T^{\mu\nu},
\ee
leading to the field equation
\be
G^{\mu\nu} = -\frac{\kappa}{2}\, T^{\mu\nu}.
\label{Eq-20}
\ee
This can, however, not yet be the final equation,
but only an approximation for weak fields. Indeed, the
identity (\ref{Eq-18}) implies $\p_\nu T^{\mu\nu} = 0$
which is unacceptable (in contrast to the charge conservation of ED).
For instance, the motion of a fluid would then not
at all be affected by the gravitational field. Clearly,
we must introduce a {\it back-reaction} on matter. Why
not just add to $T^{\mu\nu}$ in (\ref{Eq-20}) the energy-momentum
tensor $^{(2)}t^{\mu\nu}$ which corresponds to the Pauli-Fierz
Lagrangian (\ref{Eq-15})? But this modified equation cannot
be derived from a Lagrangian and is still not consistent,
but only the second step of an iteration
process:
\bdm
{\cal L}^{free} \longrightarrow ^{(2)}t^{\mu\nu} \longrightarrow {\cal L}^{cubic} \longrightarrow ^{(3)}t^{\mu\nu} \longrightarrow ... \, ?
\edm
The sequence of arrows has the following meaning:

A Lagrangian which gives the quadratic terms $^{(2)}t^{\mu\nu}$
in
\be
G^{\mu\nu} = -\frac{\kappa}{2} \left( T^{\mu\nu} + ^{(2)}t^{\mu\nu} + ^{(3)}t^{\mu\nu} + ... \right)
\label{Eq-21}
\ee
must be cubic in $h_{\mu\nu}$, and in turn leads to cubic
terms $^{(3)}t^{\mu\nu}$ of the gravitational energy-momentum
tensor. To produce these in the field equation (\ref{Eq-21}),
we need quartic terms in $h_{\mu\nu}$, etc. This is an
infinite process. By a clever reorganization it stops
already after the second step, and one arrives
at field equations which are equivalent to Einstein's
equations (S. Deser, \cite{5}). The physical metric of GR is
given in terms of
$\phi^{\mu\nu} = h^{\mu\nu} - \frac{1}{2} \eta^{\mu\nu} h$ by
\be
\sqrt{-g}\, g^{\mu\nu} = \eta^{\mu\nu} - \phi^{\mu\nu}, \;\;\;
g:=\det( g_{\mu\nu} ).
\label{Eq-22}
\ee

At this point one can reinterprete the theory
geometrically. Thereby the flat metric disappears completely
and one arrives in a pedestrian way at GR.

It should, however, be pointed out that, as a result
of gauge invariance, the energy arguments in this
reasoning are somewhat ambiguous. In view of this
we shall later (section \ref{Sect-2.5}) discuss another approach.

Let us first pursue the approximate theory, keeping
only $^{(2)}t^{\mu\nu}$ in (\ref{Eq-21}). The linearized
Bianchi identity (\ref{Eq-18}) implies the conservation laws
\be
\p_\nu \left( T^{\mu\nu} + ^{(2)}t^{\mu\nu} \right) = 0.
\label{Eq-23}
\ee
This gives
\be
\p_\nu T_\mu^{\;\nu} - \frac{\kappa}{2}\, \p_\mu h_{\alpha\beta} T^{\alpha\beta} = 0.
\label{Eq-24}
\ee
This is analogous to $\p_\nu T_\mu^{\;\nu} - F_{\mu\nu} j^\nu = 0$
in ED. For a charged test particle one obtains from this the
Lorentz equation of motion:
$\frac{\diff}{\diff\tau} (m u^\mu) = e F^{\mu}_{\;\nu} u^\nu$.
Similarly, from (\ref{Eq-24}) one can derive the following equation
of motion for a neutral test particle in a gravitational
field $h_{\alpha\beta}$:
\be
\frac{\diff u_\mu}{\diff\tau} + \kappa
\left( \p_\beta h_{\mu\alpha} - \frac{1}{2} \p_\mu h_{\alpha\beta} \right) u^\alpha u^\beta = 0.
\label{Eq-25}
\ee
Geometrically, this is just the linearization of the
geodesic equation for the metric
\be
g_{\mu\nu} = \eta_{\mu\nu} + \kappa h_{\mu\nu}
\label{Eq-26}
\ee
(compare this with (\ref{Eq-22})).
We shall soon see that this has not only a formal meaning.

\subsection{Further discussion of the linearized theory}

The linearization (equations (\ref{Eq-20}) and (\ref{Eq-25})) is,
by the way, already contained in Einstein's Zurich note book from 1912!
This is now published in Volume 3 of the {\it Collected Papers}.

It is convenient to introduce the fields
\be
\phi_{\mu\nu} = h_{\mu\nu} - \frac{1}{2} \eta_{\mu\nu} h
\label{Eq-27}
\ee
and impose the Lorentz-Hilbert gauge condition
\be
\p_\nu \phi^{\mu\nu} = 0.
\label{Eq-28}
\ee
(This is also contained in the 1912 notes of Einstein.)
Then the field equations (\ref{Eq-20}), with expression (\ref{Eq-16}),
become simply
\be
\Box\phi^{\mu\nu} = -\kappa T^{\mu\nu}.
\label{Eq-29}
\ee
This is again very similar to what we are used to in ED.
The retarded integral of the source $T^{\mu\nu}$ describes the
emission of gravitational radiation and gives, for instance,
the correct damping of binary pulsars.

In the {\it almost Newtonian limit} we have $T_{00} \approx \rho$,
and all other components are much smaller. Then only $\phi_{00}$
survives:
\bdm
\phi_{00}(\bbx) = -\frac{\kappa}{4\pi} \int \frac{ \rho(\bbx') }{ |\bbx - \bbx'| } \diff^3 x'.
\edm
Using also the Newtonian limit of the equation of
motion (\ref{Eq-25}), one finds that
\be
\kappa\phi_{00} = 4 U, \;\;\; U: \hbox{Newtonian potential},
\label{Eq-30}
\ee
and
\be
\kappa^2 = 16\pi G.
\label{Eq-31}
\ee
Therefore,
\be
\kappa h_{\mu\nu} = 2 U \delta_{\mu\nu}\, .
\label{Eq-32}
\ee
In the next order (in $1/c$) we encounter for rotating
sources the gravitomagnetic field. If the spatial stress
$T_{ij}$ can be neglected, the field equations (\ref{Eq-29})
reduce to
\bdm
\Box\phi_{ij} = 0, \;\;\;
\Box\phi_{0\mu} = -\kappa T_{0\mu}\, .
\edm
Thus $A_\mu := -\frac{1}{4}\kappa \phi_{0\mu}$
satisfy Maxwell type equations:
\bdm
\Box A_\mu = J_\mu, \;\;\;
\p^\mu A_\mu = 0,
\edm
where $J_\mu = 4\pi G T_{0\mu}$ is proportional to the mass-energy
current density. (Note that $A_0 = -U$.) It is natural to
define ``gravitational electric and magnetic fields'' $\bbE$ and
$\bbB$ by the same formulas in terms of $A_\mu$ as in ED.

Let us now assume that the time derivatives of $\phi_{\mu\nu}$ can
be neglected (quasi-stationary situations). Then $\Delta\phi_{ij}=0$
in all space, whence $\phi_{ij}=0$, and hence $A_\mu$ describes
the gravitational field. In this approximation, the equation
of motion (\ref{Eq-25}) reduces for non-relativistic velocities to
\be
\ddot{\bbx} = \bbE + 4\dot{\bbx} \wedge \bbB.
\label{Eq-33}
\ee
The factor $4$ in the ``magnetic term'' reflects the spin-$2$
character of the gravitational field. The potentials are given by
\be
A_0 = -U, \;\;\;
A_i(\bbx) = G \int \frac{ T_{0i}(\bbx') }{ |\bbx - \bbx'| } \diff^3 x'.
\label{Eq-34}
\ee
On the basis of this one obtains immediately the Lense-Thirring
precession of a gyroscope, by simply translating the spin precession
formula in electrodynamics:
Substitute in the well-known formulae
\bdm
\dot{\bbS} = \bbmu \wedge \bbB, \;\;\;
\bbmu = \frac{e}{2m} \bbS
\edm 
of ED $e$ by $m$ and $\bbB$ by $4\bbB$.
This gives the Lense-Thirring precession frequency
\be
\bbomega_{LT} = -2\curl\bbA.
\ee
The experiment Gravity Probe - B is supposed to measure this
effect directly. Its launch is scheduled for early 2001 \cite{12}.

Next, we look at the {\it coupling to the electromagnetic field}:
\bdm
{\cal L}_{int} = -\frac{\kappa}{2} h_{\mu\nu} T^{\mu\nu}_{elm}.
\edm
Maxwell's equations in the presence of the gravitational field
$h_{\mu\nu}$ follow from the Lagrangian
\bea
{\cal L}_{elm} &=& -\frac{1}{4} F_{\mu\nu} F^{\mu\nu} - j^\mu A_\mu - \frac{\kappa}{2} h_{\mu\nu} T^{\mu\nu}_{elm} \nonumber\\
               &=& -\frac{1}{4} \left( 1 + \frac{\kappa}{2} h \right) \left( \eta^{\mu\nu} - \kappa h^{\mu\nu} \right)
                     \left( \eta^{\rho\sigma} - \kappa h^{\rho\sigma} \right) F_{\mu\rho} F_{\nu\sigma} - j^\mu A_\mu \nonumber\\
               &=& -\frac{1}{4} \sqrt{-g}\, g^{\mu\nu} g^{\rho\sigma} F_{\mu\rho} F_{\nu\sigma} - j^\mu A_\mu .
\label{Eq-35}
\eea
In the last equality sign we used the metric (\ref{Eq-26}). (All
equality signs are meant to hold in lowest order of $h_{\mu\nu}$.)
This gives the modified Maxwell equations:
\be
\p_\sigma \left\{ \left( 1 + \frac{\kappa}{2} h \right)\left( \eta^{\mu\nu} - \kappa h^{\mu\nu} \right)
                  \left( \eta^{\rho\sigma} - \kappa h^{\rho\sigma} \right) F_{\mu\rho} \right\} = -j^\nu.
\label{Eq-36}
\ee
Expanding this for a diagonal $h_{\mu\nu}$ (as in (\ref{Eq-32})),
one finds the standard form of Maxwell's equations for macroscopic media,
with a dielectric constant $\epsilon$ and a magnetic permeability $\mu$.
For an almost Newtonian situation (equation (\ref{Eq-32})) these are given by
\be
\epsilon = \mu = 1 - 2U.
\label{Eq-37}
\ee
The corresponding refraction index is
\be
n = \sqrt{\epsilon\mu} = 1 - 2U.
\label{Eq-38}
\ee
This result implies the correct light deflection, and more
generally ``all'' of gravitational-lensing theory.

P. Schneider discussed in his talk some major
applications of gravitational lensing. During the past
$\sim 15$ years this field has rapidly reached an important
position in present day astronomy and astrophysics.
An exciting new result is the first evidence for
gravitational lensing by large-scale structures \cite{13}.
This has demonstrated the technical feasibility of using
weak lensing surveys to restrict the cosmological parameters.
With upcoming wide field CCD cameras much progress can
be expected.

Another very interesting recent result is the measurement
of galaxy-galaxy weak lensing from Sloan commissioning
data which show that galaxy halos are very extended, so
much that the assignment to individual galaxies becomes at
some distance meaningless \cite{14}.

\subsection{The renormalized physical metric}

Now I come to a conceptually important point: we shall
see that the flat Minkowski metric is not observable.

In order to see this we study the behavior of measuring
sticks and clocks in a gravitational field. To be specific,
we use the hydrogen atom for defining units of length and time.
Moreover, we put the H-atom into the gravitational field
outside of a spherically symmetric mass distribution of
total mass $M$ at the distance $R$ from the center, where
\be
\kappa h_{\mu\nu} = -\frac{2GM}{R} \delta_{\mu\nu}\, .
\label{Eq-39}
\ee
In this gravitational field Maxwell's equations (\ref{Eq-36})
imply the following modified Laplace-Poisson equation for the
scalar potential $\varphi$ for the proton:
\bdm
\left( 1 + \frac{2GM}{R} \right) \Delta\varphi = -e\,\delta^3 (\bbx),
\edm
whence
\be
\varphi = \frac{e}{4\pi\epsilon} \frac{1}{r}, \;\;\;
\epsilon = \frac{1}{1 - 2GM/R} .
\label{Eq-40}
\ee
This is the Coulomb potential for the effective charge
\be
e_{eff} = \frac{e}{\epsilon} .
\label{Eq-41}
\ee
The equation of motion (\ref{Eq-25}) or (\ref{Eq-33}) reduces
to the Newtonian equation
\be
m_{eff} \ddot{\bbx} = e_{eff} \frac{\bbx}{4\pi r^3},
\label{Eq-42}
\ee
with
\be
m_{eff} = m \left( 1 + 3\frac{GM}{R} \right).
\label{Eq-43}
\ee
These effective quantities determine the {\it Bohr radius}
\be
a_0 = \frac{\hbar^2}{m_{eff} (e_{eff}^2/4\pi)} = \frac{\hbar^2}{m e^2/4\pi} \left(1 - \frac{GM}{R} \right)
\label{Eq-44}
\ee
and the {\it Rydberg frequency}
\be
\omega = \frac{1}{2\hbar} m_{eff} \left( \frac{ e_{eff}^2 }{4\pi\hbar} \right)^2 
       = \frac{1}{2\hbar} m \left( \frac{ e^2 }{4\pi\hbar} \right)^2 \left(1 - \frac{GM}{R} \right).
\label{Eq-45}
\ee
Thus, using a somewhat unphysical language, we would
conclude that the atoms become smaller and frequencies
(times) decrease (increase).

It is, however, clearly much more physical, to express
this as follows: We maintain that the Bohr radius and
the Rydberg frequency define always our units of length
and time. This means that we have to rescale the
original ``unrenormalized'' length ($r$) and time ($t$) in
a spacetime dependent manner:
\be
\tilde{r} = r(1 + GM/R), \;\;\;
\tilde{t} = t(1 - GM/R).
\label{Eq-46}
\ee
In other words, the {\it physical metric} is
\be
\diff\tilde{s}^2 = \eta_{\mu\nu} \diff\tilde{x}^\mu \diff\tilde{x}^\nu
                 = g_{\mu\nu} \diff x^\mu \diff x^\nu,
\label{Eq-47}
\ee
with
\be
g_{\mu\nu} = \eta_{\mu\nu} + \kappa h_{\mu\nu}\, .
\label{Eq-48}
\ee
Clearly, this is a dynamical field. Let me stress that
in the ``renormalized'' description the speed of light is
always $1$, while it would be spacetime dependent if
we would maintain the fiction of a flat Minkowski metric.

\begin{figure}
  \begin{center}
    \includegraphics[height=0.2\textheight]{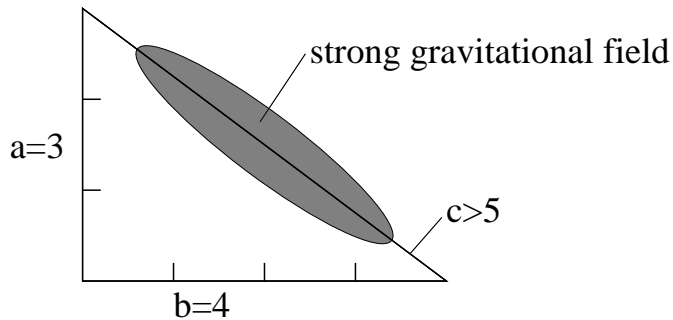}
    \caption{The failure of Pythagoras' theorem in the presence of a gravitational field.}
    \label{Fig-1}
  \end{center}
\end{figure}

In this sense spacetime is really curved. I illustrate
this in Figure \ref{Fig-1}, which demonstrates the failure
of Pythagoras' theorem.
\\

Let me summarize this discussion:

The consequent development of the theory finally made it
possible to eliminate the flat Minkowski metric, leading
to a description in terms of a curved metric which has a
direct physical meaning. The originally postulated Lorentz
invariance turned out to be physically meaningless
and plays no useful role. The flat Minkowski spacetime
becomes a kind of unobservable ether. The conclusion
is inevitable that spacetime is a pseudo-Riemannian
(Lorentzian) manifold, whereby the metric is a dynamical
field, subjected to field equations.

Once this geometrical, Einsteinian point of view
is accepted, the field equations are practically unique
(up to the cosmological term). For instance, the
vacuum field equation of Einstein is the only one
for which no derivatives of $g_{\mu\nu}$ higher than second
order are allowed (Lovelock theorem). The only
freedom is the cosmological term, to which we shall
return later.

With this fundamental step we encounter qualitatively
new phenomena. Two of the most dramatic ones are:
\begin{enumerate}
\item[(i)]{The appearance of spacetime horizons, in particular
for black hole solutions. {\it Direct} observational evidence
for such objects would clearly be most important. The
prospects for this look good. In the long run the most
important results will come from gravitational wave
astronomy, in particular from LISA.}
\item[(ii)]{Since we loose translational invariance of special
relativistic field theories, energy-momentum conservation
breaks down. This happens in a most dramatic way
in inflationary cosmological models. Only for isolated
systems we can still define total energy and momentum.}
\end{enumerate}

\subsection{Uniqueness of nonlinearities and related issues}
\label{Sect-2.5}

Before discussing some further important issues, I
list the pros and cons of the field theoretic approach.\\
\\
{\bf Advantages:}
\begin{enumerate}
\item[(i)]{This follows the pattern of well understood field
theories (e.g., electrodynamics).}
\item[(ii)]{It directly predicts that nonrelativistic gravitation
is attractive. [How this comes out in Einstein's geometric
approach will be explained shortly.]}
\item[(iii)]{It leads unavoidably to a curved spacetime structure.}
\end{enumerate}
Note, that the equivalence principle is {it not} used as a
cornerstone.\\
\\
{\bf Disadvantages:}
\begin{enumerate}
\item[(i)]{The uniqueness of the nonlinearities is not so clear,
while this is one of the most beautiful and convincing aspects
of the geometric approach. I shall say more on this below.}
\item[(ii)]{The starting point is somewhat unphysical:
Minkowski spacetime is assumed to be a good approximation.
While this is the case in most applications, it is certainly
not true close to black holes.}
\end{enumerate}
Two comments must be added to this.
\begin{enumerate}
\item{What fixes the {\it sign} on the right hand side of
Einstein's field equations
\bdm
G_{\mu\nu} = 8\pi G T_{\mu\nu} ?
\edm
{\bf Answer:} Only for the correct sign do we have a positive
energy theorem (PET). The latter says (in an untechnical language):

{\it One cannot construct an object out of ``ordinary'' matter,
i.e., matter with positive local energy density, whose total
energy (including gravitational contributions) is negative.}

This theorem is far from obvious. (A simplified proof
was given by Witten \cite{15}, making use of spinors.)
To emphasize its significance, I note the following:
\begin{enumerate}
\item{If objects with negative total energy (mass) could exist
in GR, they would {\it repel} rather than attract nearby objects}
\item{The PET implies, for example, that there are no regular
interior solutions for the ``anti-Schwarzschild'' vacuum solution
with negative mass, and hence there is attraction.}
\end{enumerate}

This is only true for the correct sign of Einstein's field
equation. Changing the sign would lead to repulsion.
Since the PET would, however, no more be true, we
could presumably extract an {\it unlimited} amount of
energy from a system with negative energy.

With the correct sign this bizarre situation cannot
occur in GR. {\it Physically}, the reason is that as a
system is compressed to take advantage of the negative
gravitational binding energy, a {\it black hole} is inevitably
formed which has {\it positive total energy}.
}
\item{{\bf Perturbation consistency and uniqueness}\\
The uniqueness of Einstein's vacuum equation $G_{\mu\nu}[\bbg]=0$
(ignoring the $\Lambda$-term) can be translated into a perturbative
consistency property, which we conversely may impose in
the field theoretic approach to guarantee uniqueness. Let
me explain this in some detail.

First, we decompose the Einstein tensor as
\be
G_{\mu\nu} = G_{\mu\nu}^{(1)} + G_{\mu\nu}^{(2)} + ...,
\label{Eq-49}
\ee
where $G_{\mu\nu}^{(n)}$ contains all terms of power $n$ in the field
variable $g_{\mu\nu}$. For the metric we make a perturbative (formal)
expansion about the Minkowski metric
\be
g_{\mu\nu} = \eta_{\mu\nu} + \epsilon g_{\mu\nu}^{(1)} + \epsilon^2 g_{\mu\nu}^{(2)} + ... .
\label{Eq-50}
\ee
Inserting this into the vacuum equation leads to an
infinite chain of equations:
\bea
&& G_{\mu\nu}^{(1)}[ \bbg^{(1)} ] = 0, \nonumber\\
&& G_{\mu\nu}^{(2)}[ \bbg^{(1)} ] + G_{\mu\nu}^{(1)}[ \bbg^{(2)} ] = 0, \label{Eq-51}\\
&& ... \, . \nonumber
\eea
Now we take the divergence of the second equation and use
the linearized Bianchi identity
\be
\p^\nu G_{\mu\nu}^{(1)} \equiv 0.
\label{Eq-52}
\ee
This gives
\be
\p^\nu G_{\mu\nu}^{(2)}[ \bbg^{(1)} ] = 0,
\label{Eq-53}
\ee
and looks like an additional requirement for $g_{\alpha\beta}^{(1)}$,
besides the linearized Einstein equation (first equation in (\ref{Eq-51})).
That would be dangerous. However, equation (\ref{Eq-53}) is {\it automatically}
satisfied because of the Bianchi identity in second order.
The same happens in higher order. Recall in this context,
that the Bianchi identity can be regarded as a consequence
of gauge invariance (general coordinate invariance).

The idea is now to turn the argument around. The
question is: If we impose the correct linearized Einstein
equations plus perturbative consistency, do the Einstein
equations uniquely follow, if no derivatives higher than
second order are allowed?

R. Wald \cite{6} has analyzed this question, and came up
with a qualified ``yes''. There are still some loose ends,
but these are presumably not serious. T. Damour informed me at
this meeting, that in collaboration with M. Henneaux the
remaining gaps have been closed.

So general invariance is unavoidable, at least for
reasonable matter couplings\footnote{There exist nonlinear {\it vacuum}
theories with normal spin-$2$ gauge invariance:
$h_{\mu\nu} \longrightarrow h_{\mu\nu} + \p_\mu\xi_\nu + \p_\nu\xi_\mu$.
An example is
\bdm
{\cal L} = {\cal L}^{(2)} + {\cal L}^{(3)}, \;\;\;
{\cal L}^{(3)} = \left( R^{(1)} \right)^3,
\edm
where $R^{(1)}$ is the (gauge-invariant) linearized Ricci scalar.
Explicitly,
\bdm
{\cal L}^{(3)} = \left( \p^\mu\p^\nu h_{\mu\nu} - \Box h \right)^3.
\edm}.

All this reflects once more the rigidity and beauty
of GR. This theory must be contained in the low energy
limit of any true unification.
}
\end{enumerate}

Point 2. is also relevant for string-theory: The string
excitations contain a massless spin-$2$ mode, and
therefore GR has to be part of the field theoretic limit
of string theory.

\section{Tensor-scalar generalizations}
\label{Sect-3}

In the light of the marvellous rigidity of GR and the many
tests it has already passed \cite{16}, there seem to be no good
reasons for studying alternative theories of gravity.
There is, however, one class of generalizations which
not only has a long tradition, but also new motivation
from string theory.

Already in his geometric five-dimensional unification
of gravity and electromagnetism, Kaluza \cite{17} automatically got
also a scalar component for the $4$-dimensional theory.
Indeed, the appearance of scalar fields is unavoidable
in Kaluza-Klein theories. Later, Jordan \cite{18} tried to make
use of the scalar field to obtain a theory in which the
gravitational constant is replaced by a dynamical field.
This work was criticized by Fierz \cite{19}, who noted that Jordan's
tensor-scalar theories generically entail unacceptable
violations of the equivalence principle. Fierz specialized
Jordan's theory such that this was avoided and arrived
at a theory which was later called the Brans-Dicke-theory.
(More on this, as well as references, can be found in \cite{20}.)

This is a special case of a class of metrically-coupled
tensor-scalar theories which can be characterized by the
following two postulates:
\begin{enumerate}
\item[(i)]{ {\bf Metric coupling to matter:} Only the ``physical''
metric (not the scalar field) couples directly to matter,
as in GR: ${\cal L}_{mat} [\psi; g_{\mu\nu} ]$
$(\p_\mu \rightarrow \nabla_\mu)$.}
\item[(ii)]{ {\bf Dynamics of $g_{\mu\nu}$, $\varphi$}:
This is given by the Einstein action for a conformally related metric,
$\tilde{g}_{\mu\nu} = f^2(\varphi) g_{\mu\nu}\,$, plus the kinetic
term for $\varphi$:
\bdm
S_{grav} = -\frac{1}{16\pi G} \int
\left[ R[\tilde{g}] - 2\tilde{g}^{\mu\nu} \p_\mu\varphi\, \p_\nu\varphi \right]
\sqrt{-\tilde{g}}\,\, \diff^4 x.
\edm
}
\end{enumerate}
In this class of theories the function $f(\varphi)$ is arbitrary.
[ For the Fierz-Brans-Dicke theory $f(\varphi) = e^{\alpha\varphi}$. ]

The observable consequences of these theories have been
worked out, in particular, by T. Damour and collaborators
(see, e.g., \cite{21}). There is the interesting
possibility that the deviations from GR in weak field
situations are tiny, but become significant in strong field
regions such as the interiors of neutron stars.

I should add that in string theory scalar fields,
notably the dilaton field, appear as necessary partners
of the metric field. In some scenarios such scalar fields
behave in laboratory and solar system measurements as
massless fields, and could modify the predictions of GR.
In particular, violations of the weak equivalence
principle are expected at some level. In the light
of this, the STEP experiment is of importance.
In string theories, scalar fields typically have
couplings not much weaker than gravity. If the test
of the equivalence principle is improved by a factor
$10^6$, this would put a severe restriction on models.

It is, however, more likely that scalar fields are
{\it massive}, in which case the theory is practically
equivalent to GR.

Generalized tensor-scalar theories are often used
in cosmological model building (inflation, quintessence,
etc). Much of this is, however, quite arbitrary and
very speculative.

Finally, I should mention that scalar-tensor
gravitational waves can have an additional transverse
breathing mode. The strength of this mode depends,
of course, on the nature of the source.\\
\\

\section{The mystery of the cosmic vacuum energy density}

{\it Classically}, one may ignore the cosmological term in
Einstein's field equation, although there is no good reason
for this, since it is allowed by the principles
of GR. (Simplicity is not a convincing argument.)

In {\it quantum theory} the $\Lambda$-problem is much worse,
because quantum fluctuations are expected to give rise
to a non-vanishing vacuum energy density $\rho_{vac}$, which acts
like a cosmological constant.
Without gravity, we do, of course, not care about the
energy of the vacuum, because only energy {\it differences}
matter. However, even then the quantum fluctuations of
the vacuum can be important, as is beautifully
demonstrated by the Casimir effect. The radiative
corrections to Maxwell's equations, first discussed
by Heisenberg and Euler, and later by Weisskopf,
can also be interpreted in this manner (see, e.g., \cite{22}).

When we consider the coupling to gravity, the
vacuum expectation value of the energy-momentum
tensor has the form of the cosmological term (up to
higher order curvature contributions):
\bdm
\langle T_{\mu\nu} \rangle_{vac} = g_{\mu\nu}\, \rho_{vac} + ... \, .
\edm
The {\it effective} cosmological constant, which controls
the large scale  behavior of the universe, is given by
$\Lambda = 8\pi G \rho_{vac} + \Lambda_0$,
where $\Lambda_0$ is a bare cosmological
constant in Einstein's field equations. We know
that $\rho_{\Lambda} \equiv \Lambda/8\pi G$ can not
be much larger than the critical density,
$\rho_{crit} = 8\times 10^{-47} h_0^2 GeV^4$
($h_0 \equiv H_0/ 100 km/s/Mpc$).
This is infinitesimal by particle physics standards.

It is recognized since quite some time that
this is a profound mystery. Indeed, we expect that quantum
fluctuations in the fields of the standard model of
particle physics, cut off at about the Fermi scale, contribute
to the vacuum energy density, because there is no
symmetry principle in this energy range which would
require a cancellation of the various contributions
(as in strictly supersymmetric theories).

To have some measure, let us compare $\rho_{crit}$ with
the condensation energy density of QCD in the broken phase
of the chiral symmetry, which is about
$\Lambda^4_{QCD}/16\pi^2 \approx 10^{-4} GeV^4$.
The discrepancy is at least $40$ orders of magnitudes.

So far string theory has not offered convincing clues
why the cosmological constant is extremely small (for
a recent discussion, see \cite{23}). The main reason
is that a low energy mechanism is required, and the
low energy physics is described by the standard model.

\section{Quintessence}

G. Tammann reviewed the recent astronomical evidence
for a cosmologically significant vacuum energy density
(or some effective equivalent). This arises mainly from
the Hubble diagram of type Ia supernovae and from
the observed temperature fluctuations of the cosmic
microwave background radiation. In particular, the first
results from the BOOMERANG experiment have reinforced
the evidence.

If the present situation is going to stay we are
confronted with the following {\it cosmic coincidence problem:}
Since the vacuum energy is constant in time, while the matter
energy density decreases as the universe expands, it is
more than surprising that the two are comparable just at
the present time, while their ratio has been tiny in
the early universe.

Possible ways of avoiding this puzzle have recently
been discussed extensively. The general idea is to explain the
accelerated expansion of the universe by yet another form
of exotic missing energy with negative pressure, called
{\it quintessence}. In concrete models this is described by
a scalar field, whose dynamics is such that its energy
naturally adjusts itself to be comparable to the matter
density today for generic initial conditions.

Let me briefly describe a simple model of this kind \cite{24}.
For the dynamics of the scalar field $\phi$ we adopt an
exponential potential
\bdm
V = V_0\, e^{-\lambda\phi/M_P}.
\edm
Such potentials often arise in Kaluza-Klein and string
theories. Matter is described by a fluid with a baryotropic
equation of state: $p_f = (\gamma-1) \rho_f$.

For a Friedmann model with zero space-curvature, one
can cast the basic equations into an autonomous two-dimensional
dynamical system for the quantities
\bdm
x(\tau) = \frac{\kappa\dot{\phi}}{\sqrt{6} H}, \;\;\;
y(\tau) = \frac{\kappa\sqrt{V}}{\sqrt{3} H},
\edm
where
\bdm
H = \dot{a}/a, \;\;\;
\tau = \log a, \;\;\;
\kappa^2 = 8\pi G
\edm
($a(t)$ is the scalar factor). This system of autonomous
differential equations has the form
\bdm
\frac{\diff x}{\diff\tau} = f(x,y; \lambda,\gamma), \;\;\;
\frac{\diff y}{\diff\tau} = g(x,y; \lambda,\gamma),
\edm
where $f$ and $g$ are polynomials in $x$ and $y$ of third
degree, which depend parametrically on $\lambda$ and $\gamma$.
The density parameters $\Omega_\phi$ and $\Omega_f$ for the
field $\phi$ and the fluid are given by
\bdm
\Omega_\phi = x^2 + y^2, \;\;\;
\Omega_\phi + \Omega_f = 1.
\edm

\begin{figure}
  \begin{center}
    \includegraphics[height=0.4\textheight]{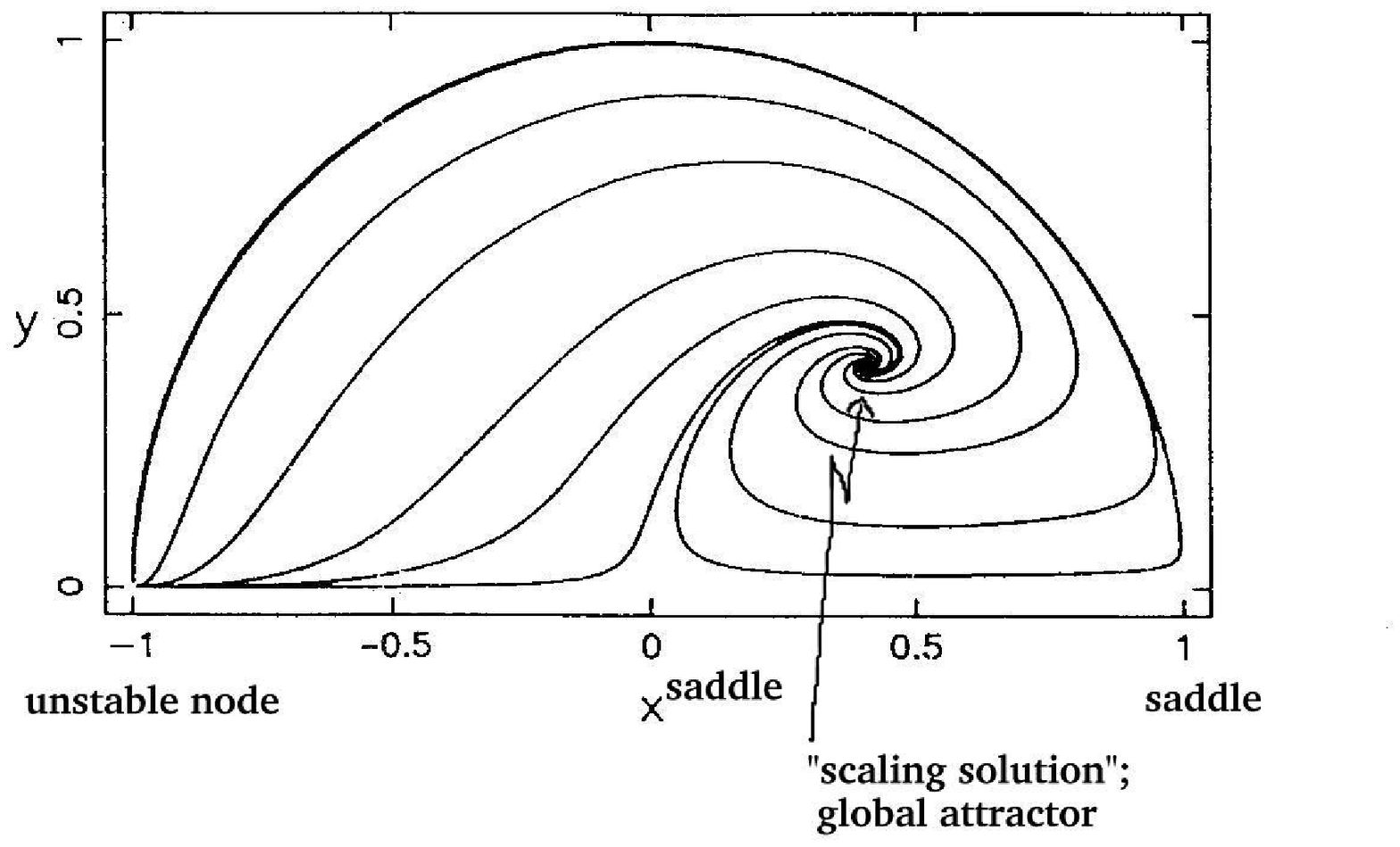}
    \caption{Phase plane for $\gamma=1$, $\lambda=3$. The late-time attractor is the scaling solution with $x=y=1/\sqrt{6}$ (from Ref. \cite{24}).}
    \label{Fig-2}
  \end{center}
\end{figure}

The interesting fact is that, for a large domain of the
parameters $\lambda$, $\gamma$, the phase portrait has
qualitatively the shape of Figure \ref{Fig-2}. Therefore, under
generic initial conditions, there is a global attractor
(a node or a spiral) for which $\Omega_\phi = 3\gamma/\lambda^2$.
For this ``scaling solution'' $\Omega_\phi/\Omega_f$
remains fixed, and for any other solution this ration is
finally approached.

This looks good. However, various complications of
the model introduce also unstable directions and the
attracting behavior gets lost. Moreover, if a
constant of order $M_{Fermi}^4$ (or even $m_e^4$) would
be added to the potential $V$, the mechanism would not work.
In addition, we have to worry about unacceptable changes in
the nucleosynthesis results.\\
\begin{center}
* \quad * \quad *
\end{center}

Having expressed once more that we are confronted with
a deep mystery, I conclude with the following amusing
story:

During the 1920ties most people were convinced that
the universe is on the average {\it static}. The
ground-breaking papers of Friedmann and Lema\^{\i}tre
were, in fact, largely ignored. In comments to Lema\^{\i}tre
during the Solvay Meeting in 1927, Einstein rejected the
expanding universe solutions as physically acceptable.
It is also not well-known that Hubble interpreted his
famous results on the redshift of the radiation emitted
by distant nebulae in the framework of the static de Sitter
model. However, Lema\^{\i}tre's successful explanation
of Hubble's discovery finally changed the viewpoint of
the majority of workers in the field. At this point Einstein
rejected the cosmological term as superfluous and no
longer justified. He published his new view in
the ``Sitzungsberichte der Preussischen Akademie der
Wissenschaften''. The correct citation is:\\

Einstein, A. (1931). Sitzungsber. Preuss. Akad. Wiss. 235-37.\\

Many people have quoted this paper, but never read it.
As a result, the quotations gradually changed in an
interesting, quite systematic fashion. Some steps are
shown in the following sequence:
\begin{enumerate}
\item[-]{A. Einstein. 1931. Sitzsber. Preuss. Akad. Wiss. ...}
\item[-]{A. Einstein. Sitzber. Preuss. Akad. Wiss. ... (1931)}
\item[-]{A. Einstein (1931). Sber. Preuss. Akad. Wiss. ...}
\item[-]{A. Einstein. .. 1931. Sb. Preuss. Akad. Wis. ...}
\item[-]{A. Einstein. S.-B. Preuss. Akad. Wis. ... 1931}
\item[-]{A. Einstein. S.B. Preuss. Akad. Wiss. (1931) ...}
\item[-]{A. Einstein. and Preuss. S. B. (1931) Akad. Wiss. ...}
\end{enumerate}

Presumably, one day some historian of science will
try to find out what happened with the young physicist
S.B. Preuss, who apparently wrote just one paper and
then disappeared from the scene.\\

With this light note I would like to conclude.


\begin{thebibliography}{10}

\newcommand{\AJM}{{\it Am.\ J.\ Math.\ }}
\newcommand{\AM} {{\it Ann.\ Math.\ }}
\newcommand{\AMC}{{\it Ann.\ Math.\ Soc.\ Coll.\ (4th ed.)\ }}
\newcommand{\ANY}{{\it Ann.\ N.Y.\ Acad.\ Sci.\ }}
\newcommand{\AP} {{\it Ann.\ of Phys.\ }}
\newcommand{\APL}{{\it Ann.\ Phys.\ (Leipzig)\ }}
\newcommand{\ASI}{{\it American\ Scientist\ }}
\newcommand{\AIA}{{\it Ann.\ Inst.\ H.\ Poincar\'e\ A\ }}
\newcommand{\AJ} {{\it Astrophys.\ J.\ }}
\newcommand{\AaA}{{\it Astronomy and Astrophysics }}
\newcommand{\BLM}{{\it Bull.\ London\ Math.\ Soc.\ }}
\newcommand{\CQG}{{\it Class.\ Quantum\ Grav.\ }}
\newcommand{\CMP}{{\it Comm.\ Math.\ Phys.\ }}
\newcommand{\CPA}{{\it Commun.\ Pure \ Appl. \ Math.\ }}
\newcommand{\GRG}{{\it Gen.\ Rel.\ Grav.\ }}
\newcommand{\HPA}{{\it Helv.\ Phys.\ Acta\ }}
\newcommand{\IJTP}{{\it Int.\ J. \ Theo.\ Phys.\ }}
\newcommand{\JET}{{\it JETP\ Lett.\ }}
\newcommand{\JDG}{{\it J.\ Diff.\ Geom.\ }}
\newcommand{\JMP}{{\it J.\ Math.\ Phys.\ }}
\newcommand{\JPM}{{\it J.\ Phys.\ A:\ Math.\ Gen.\ }}
\newcommand{\LMP}{{\it Lett.\ Math.\ Phys.\ }}
\newcommand{\MRA}{{\it Mon.\ Not.\ R.\ Astron.\ Soc.\ }}
\newcommand{\MPLA}{{\it Mod.\ Phys.\ Lett.\ A\ }}
\newcommand{\NAT}{{\it Nature\ }}
\newcommand{\NATL}{{\it Nature\ (London)\ }}
\newcommand{\NPS}{{\it Nature\ (Phys.\ Sci.)\ }}
\newcommand{\NC} {{\it Nuovo\ Cimento\ }}
\newcommand{\NP} {{\it Nucl.\ Phys.\ }}
\newcommand{\NPB}{{\it Nucl.\ Phys.\ B\ }}
\newcommand{\NPBP}{{\it Nucl.\ Phys.\ B\ (Proc. Suppl.)\ }}
\newcommand{\PL} {{\it Phys.\ Lett.\ }}
\newcommand{\PLA}{{\it Phys.\ Lett.\ A\ }}
\newcommand{\PLB}{{\it Phys.\ Lett.\ B\ }}
\newcommand{\PRL}{{\it Phys.\ Rev.\ Lett.\ }}
\newcommand{\PR} {{\it Phys.\ Rev.\ }}
\newcommand{\PRB}{{\it Phys.\ Rev.\ (Sect.\ B)\ }}
\newcommand{\PRP}{{\it Phys.\ Rep.\ }}
\newcommand{\PRD}{{\it Phys.\ Rev.\ D\ }}
\newcommand{\RMP}{{\it Rev.\ Mod.\ Phys.\ }}
\newcommand{\PZ}{{\it Phys.\ Z.\ }}
\newcommand{\PAW}{{\it Preuss.\ Akad.\ Wiss.\ Berlin,\ Sitz.ber.\ II\ }}

\newcommand{\PTR}{{\it Phil.\ Trans.\ R.\ Soc.\ (London)\ }}
\newcommand{\SDA}{{\it Sitzungsber.\ K.\ Preuss.\ Akad.\ Wiss.\ Phys.\ Math.\ Kl.\ }}
\newcommand{\SLA}{{\it Proc.\ Roy.\ Soc.\ (London)\ Ser.\ A\ }}
\newcommand{\PNA}{{\it Proc.\ Natl.\ Acad.\ Sci.\ }}
\newcommand{\PKN}{{\it Proc.\ Kon.\ Ned.\ Akad.\ Wet.\ }}
\newcommand{\PIA}{{\it Proc.\ Roy.\ Irish\ Acad.\ }}
\newcommand{\PTP}{{\it Progr.\ Theor.\ Phys.\ (Kyoto)\ }}
\newcommand{\RPP}{{\it Rep.\ Prog.\  Phys.\ }}
\newcommand{\RNC}{{\it Riv.\ Nuovo\ Cimento\ }}
\newcommand{\SJN}{{\it Sov.\ J.\ Nucl.\ Phys.\ }}
\newcommand{\TSM}{{\it Trans.\ Am.\ Math.\ Soc.\ }}
\newcommand{\ZPC}{{\it Z.\ Phys.\ C\ }}


\bibitem{1}
R.P. Feynman, F.B. Morinigo, and W.G. Wagner,
{\em Feynman Lectures on Gravitation},
edited by Brian Hatfield (Addison-Wesley, Reading, 1995).

\bibitem{2}
W. Wyss,
\HPA {\bf 38}, 469 (1965).

\bibitem{3}
W. Thirring,
\AP {\bf 16}, 96 (1961).

\bibitem{4}
S. Weinberg,
\PR {\bf 138}, 988 (1965).

\bibitem{5}
S. Deser,
\GRG {\bf 1}, 9 (1970).

\bibitem{6}
R.M. Wald,
\PRD {\bf 33}, 3613 (1986).

\bibitem{7}
G. Nordstr\"om,
\PZ {\bf 13}, 1126 (1912);
\APL {\bf 40}, 856 (1913);
\APL {\bf 42}, 533 (1913).

\bibitem{8}
A. Einstein and A.D. Fokker,
\APL {\bf 44}, 321 (1914).

\bibitem{9}
J. Ehlers and W. Rindler,
\GRG {\bf 29}, 519 (1997).

\bibitem{10}
W. Pauli and M. Fierz,
\HPA {\bf 12}, 297 (1939);
\SLA {\bf 173}, 211 (1939).

\bibitem{11}
G. Wentzel,
{\em Qunatum Theory of Fields},
Interscience Publishers (1949); especially \S 22.

\bibitem{12}
For information about the project, see http://einstein.stanford.edu./.
See also the report by F. Everitt at this workshop.

\bibitem{13}
D. Bacon, A Refregier, and R. Ellis,
astro-ph/0003008;
N. Kaiser, G. Wilson, G. Luppino, and H. Dahle,
astro-ph/9907229;
L. Van Waerbeke, {\it et al.},
astro-ph/0002500.

\bibitem{14}
Ph. Fischer, {\it et al.},
astro-ph/9912119

\bibitem{15}
E. Witten,
\CMP {\bf 80}, 381 (1981).

\bibitem{16}
C. Will,
{\em The confrontation between general relativity and experiment: A 1998 update},
gr-qc/9811036.

\bibitem{17}
Th. Kaluza,
\SDA, 966 (1921).

\bibitem{18}
P. Jordan,
\NATL {\bf 164}, 637 (1949);
{\em Schwerkraft und Weltall}, 2nd ed. (Vieweg, Braunschweig, 1954).

\bibitem{19}
M. Fierz,
\HPA {\bf 29}, 128 (1956).

\bibitem{20}
L. O'Raifeartaigh and N. Straumann,
\RMP {\bf 72}, 1 (2000).

\bibitem{21}
T. Damour,
in Proceedings of the XIX th Texas Symposium on Relativistic Astrophysics and Cosmology,
\NPBP {\bf 80}, 41 (2000).

\bibitem{22}
L.D. Landau and E.M. Lifschitz,
Vol. 4 {\em Quantum Electrodynamics}, especially \S 129.

\bibitem{23}
E. Witten,
hep-th/0002297.

\bibitem{24}
E.J. Copeland, AR. Liddle, and D. Wands,
\PRD {\bf 57}, 4686 (1998).


\end{thebibliography}
\end{document}